\documentclass[12pt]{article}
\usepackage{amssymb,epsfig}

\renewcommand{\thefootnote}{\fnsymbol{footnote}}

\begin{document}

\title{
\begin{flushright}
\ \\*[-80pt] 
\begin{minipage}{0.2\linewidth}
\normalsize
arXiv:0712.2519  \\
YITP-07-90 \\
KUNS-2116 \\*[50pt]
\end{minipage}
\end{flushright}
{\Large \bf 
Metastable supersymmetry breaking vacua from 
conformal dynamics 
\\*[20pt]}}

\author{
\centerline{
Hiroyuki~Abe$^{1,}$\footnote{
E-mail address: abe@yukawa.kyoto-u.ac.jp}, \ 
Tatsuo~Kobayashi$^{2,}$\footnote{
E-mail address: kobayash@gauge.scphys.kyoto-u.ac.jp} \ and \ 
Yuji~Omura$^{3,}$\footnote{E-mail address: omura@scphys.kyoto-u.ac.jp}
}\\*[20pt]
\centerline{
\begin{minipage}{1.2\linewidth}
\begin{center}
$^1${\it \normalsize 
Yukawa Institute for Theoretical Physics, Kyoto University, 
Kyoto 606-8502, Japan} \\
$^2${\it \normalsize 
Department of Physics, Kyoto University, 
Kyoto 606-8502, Japan} \\
$^3${\it \normalsize 
Department of Physics, Kyoto University, 
Kyoto 606-8501, Japan} 
\end{center}
\end{minipage}
}\\*[50pt]}

\date{
\centerline{\small \bf Abstract}
\begin{minipage}{0.9\linewidth}
\medskip 
\medskip 
\small
We study the scenario that conformal dynamics leads to 
metastable supersymmetry breaking vacua.
At a high energy scale, the superpotential is not R-symmetric, 
and has a supersymmetric minimum.
However, conformal dynamics suppresses several operators 
along renormalization group flow toward the infrared fixed point.
Then we can find an approximately R-symmetric superpotential, 
which has a metastable supersymmetry breaking vacuum, 
and the supersymmetric vacuum moves far away from 
the metastable supersymmetry breaking vacuum.
We show a 4D simple model.
Furthermore, we can construct 5D models with 
the same behavior, because of the AdS/CFT dual. 
\end{minipage}
}

\begin{titlepage}
\maketitle
\thispagestyle{empty}
\end{titlepage}


\renewcommand{\thefootnote}{\arabic{footnote}}
\setcounter{footnote}{0}

\section{Introduction}

Conformal dynamics provides  several interesting 
aspects in supersymmetric models as well as 
non-supersymmetric models, because 
conformal dynamics exponentially suppresses or enhances 
certain operators.
For example, contact terms like $\int d^4\theta |X|^2|Q|^2$ 
are suppressed exponentially by conformal dynamics 
in the model that the chiral superfield $X$ 
belongs to the hidden conformal sector and 
the chiral superfield $Q$ belongs not to the 
conformal sector, but to the visible sector.
Such conformal suppression mechanism, i.e. conformal 
sequestering, is quite important to model building for 
supersymmetry (SUSY)
breaking \cite{Luty:2001jh,Dine:2004dv,Sundrum:2004un,Ibe:2005pj,
Schmaltz:2006qs,Murayama:2007ge}.
When $X$ contributes to SUSY breaking sizably, 
the above contact terms, in general, induce 
flavor-dependent soft SUSY breaking terms, soft sfermion masses 
and the so-called A-terms, 
and they lead to flavor changing neutral current processes, 
which are strongly constrained by current experiments.
However, conformal sequestering can suppress 
the above contact terms and flavor-dependent 
contributions to soft SUSY breaking terms.
Then, flavor-blind contributions such as 
anomaly mediation \cite{Randall:1998uk} would become dominant.

Another interesting aspect is that conformal dynamics 
can generate hierarchical structure of Yukawa couplings 
for quarks and leptons \cite{Nelson:2000sn,Kobayashi:2001kz}.
Suppose that our quark and lepton superfields $Q$ couple 
with the conformal sector like $\int d^2 \theta h Q X_1 X_2$, 
and the fields $X_1$ and $X_2$ have negative 
anomalous dimensions by conformal dynamics.
When this coupling $h$ is driven toward an infra-red (IR) 
fixed point, the fields $Q$ would have large and positive 
anomalous dimensions.
Because of such large and positive anomalous dimensions, 
Yukawa couplings among quarks/leptons Q and the 
electroweak Higgs fields become exponentially suppressed 
toward the IR direction.
Then, even if all of Yukawa couplings are 
of $O(1)$ at a high energy scale, 
hierarchies among Yukawa couplings could be 
generated by conformal dynamics.
At the same time, sfermion masses are exponentially 
suppressed toward the IR fixed point 
\cite{Nelson:2000sn,Kobayashi:2001kz,Karch:1998qa}.\footnote{
A similar dynamics would be useful to control a 
large radiative correction on Higgs soft masses \cite{Kobayashi:2004pu}.}

Here we study a new application of conformal dynamics 
for supersymmetric models, that is, 
realization of metastable SUSY breaking vacua 
by conformal dynamics.
Its idea is as follows.
The Nelson-Seiberg argument \cite{Nelson:1993nf} implies 
that 
generic superpotential has a SUSY minimum, 
but R-symmetric superpotential has no SUSY minimum, 
that is, SUSY is broken in such a model.
Thus, if we add explicit R-symmetry breaking terms 
in R-symmetric superpotential, 
a SUSY minimum would appear.
However, when such R-symmetry breaking terms are 
tiny, the previous SUSY breaking minimum would 
survive and a newly appeared SUSY preserving minimum 
would be far away from the SUSY breaking point 
in the field space.
That is the metastable SUSY breaking vacuum 
\cite{Intriligator:2007py,Shih:2007av,Abe:2007ax}.
We try to realize such a metastable SUSY 
breaking vacuum by conformal dynamics.
We start with a superpotential without R-symmetry.
However, we assume the conformal dynamics.
Because of that, certain couplings are exponentially 
suppressed.
Then, we could realize an R-symmetric superpotential 
or an approximately R-symmetric superpotential 
with tiny R-symmetry breaking terms.
It would lead to a stable or metastable SUSY breaking vacuum.
We study this scenario by using a simple model.
Also, we study 5D models, which have the same behavior.

This paper is organized as follows.
In the next section, we give a 4D simple model 
to realize our conformal scenario.
In section 3, we study 5D models, which have 
the same behavior.
Section 4 is devoted to conclusion and discussion.


\section{4D conformal model}

Our model is the $SU(N)$ gauge theory with $N_f$ flavors 
of chiral matter fields $\phi_i$ and $\tilde \phi_i$, which are 
fundamental and anti-fundamental representations of 
$SU(N)$.
The flavor number satisfies $N_f \geq \frac32 N$, 
and that corresponds to the conformal window 
\cite{Seiberg:1994pq,Intriligator:1995au},  
that is, this theory has an IR fixed point \cite{Banks:1981nn}.
The NSVZ beta-function of physical gauge coupling 
$\alpha=g^2/8\pi^2$ is  
\begin{equation}
\beta^{\rm NSVZ}_\alpha= - \frac{\alpha^2}{1-N\alpha} (3N-N_f+N_f \gamma_\phi),
\end{equation}
where $\gamma_\phi$ is the anomalous dimension of 
$\phi_i$ and $\tilde \phi_i$ \cite{Novikov:1983uc, ArkaniHamed:1997ut}.
Since the IR fixed point corresponds to $\beta^{\rm NSVZ}_\alpha=0$, 
around that point the matter fields $\phi_i$ and $\tilde \phi_i$ have 
anomalous dimensions $\gamma_\phi= -(3N-N_f)/N_f$, which 
are negative.

In addition to the fields $\phi_i$ and $\tilde \phi_i$, 
we introduce singlet fields $\Phi_{ij}$ for $i,j=1,\cdots,N_f$.
The gauge invariance allows the following superpotential 
at the renormalizable level,
\begin{equation}
W=h\phi_i\Phi_{ij}\tilde \phi_j + f {\rm Tr}_{ij }\Phi_{ij} + 
\frac{m}{2} {\rm Tr}_{ik} \Phi_{ij}\Phi_{jk} + \frac{\lambda}{3} 
{\rm Tr}_{i\ell}\Phi_{ij}\Phi_{jk}\Phi_{k\ell}.
\label{W-1}
\end{equation}
Here we have preserved the $SU(N_f)$ flavor symmetry.
Even if the $SU(N_f)$ flavor symmetry is broken, e.g 
by replacing $f {\rm Tr}_{ij }\Phi_{ij}$ by $f_{ij}\Phi_{ij}$, 
the following discussions would be valid.
For simplicity, we assume that all of couplings, 
$h$, $f$, $m$, $\lambda$, are real, although the following 
discussions are available for the model with 
complex parameters, $h$, $f$, $m$ and $\lambda$.
We can add the mass terms of $\phi_i$ and $\tilde \phi_j$ 
to the above superpotential.
We will comment on such terms later, but 
at the first stage we study the superpotential 
without the mass terms of $\phi_i$ and $\tilde \phi_j$.

If $m =\lambda =0$, the above superpotential 
corresponds to the superpotential of the 
Intriligator-Seiberg-Shih (ISS) model 
\cite{Intriligator:2006dd}.
We consider that our theory is an effective theory with 
the cutoff $\Lambda$.
We assume that dimensionless parameters $h$ and $\lambda$ 
are of $O(1)$ and dimensionful parameters 
$f$ and $m$ satisfy $f \approx m^2$ and $m \ll \Lambda$.
We denote physical couplings as 
$\hat h=(Z_\phi Z_{\tilde \phi} Z_\Phi)^{-1/2}h$, 
$\hat f_{ij} = (Z_\Phi)^{-1/2} f_{ij}$, 
$\hat m = (Z_\Phi)^{-1} m$ and $\hat \lambda = (Z_\Phi)^{-3/2} \lambda$, 
where $Z_\phi, Z_{\tilde \phi}, Z_\Phi$ are wavefunction renormalization 
constants for $\phi,\tilde \phi, \Phi$, respectively.

The F-flat conditions are obtained as 
\begin{eqnarray}
\partial_{\Phi_{ij}}W &=& h\phi_i \tilde \phi_j + f\delta_{ij} 
+ m \Phi_{ij} + \lambda \Phi_{jk}\Phi_{ki}
=0, \\
\partial_{\phi_i} W &=& h \Phi_{ij}\tilde \phi_j =0, \\
\partial_{\tilde \phi_j} W &=& h \phi_i \Phi_{ij} =0.
\end{eqnarray}
These equations have a supersymmetric solution for generic values of 
parameters, $h, m, \lambda$.
To see such a supersymmetric solution, 
following \cite{Intriligator:2006dd} we decompose $\phi, \tilde \phi$ 
and $\Phi$ as 
\begin{equation}
\Phi = \left(
\begin{array}{cc}
Y & Z^T \\
\tilde Z & X
\end{array}
\right), 
\qquad 
\phi = \left(
\begin{array}{c}
\chi \\
\rho
\end{array}
\right),
\qquad 
\tilde \phi^T = \left(
\begin{array}{c}
\tilde \chi \\
\tilde \rho
\end{array}
\right),
\end{equation}
where $Y$, $\chi$ and $\tilde \chi$ are $N\times N$ matrices, 
$X$ is an $(N_F-N)\times (N_F-N)$ matrix, 
$Z$, $\tilde Z$, $\rho$ and $\tilde \rho$ are 
$(N_F-N)\times N$ matrices.
Let us consider the slice with $Z=\tilde Z = \rho =0$ 
in the field space, where the first derivatives of $W$ 
reduce to 
\begin{equation}
W_{\Phi_{ij}} = \left(
\begin{array}{cc}
f\delta_{ij} + h \chi_i \tilde \chi_j + mY_{ji} + \lambda 
Y_{jk}Y_{ki} & 0 \\
0 & f\delta_{ij} + mX_{ji} + \lambda X_{jk}X_{ki} 
\end{array}
\right),
\end{equation}
\begin{equation}
W_{\phi_i}^T= \left(
\begin{array}{c}
hY_{ij}\tilde \chi_j \\
0
\end{array}
\right), \qquad 
W_{\tilde \phi_j} = \left(
\begin{array}{c}
h\chi_i Y_{ij} \\
0
\end{array}
\right).
\end{equation}
Here, we have used the same indices for 
$\Phi_{ij}$, $\phi_i$, $\tilde \phi_j$ and their submatrices.
Thus, the fields $X_{ij}$ and the others are decoupled 
in the F-flat conditions, $W_{\Phi_{ij}}=W_{\phi_i}=W_{\tilde
  \phi_j}=0$.
The F-flat condition $W_{\Phi_{ij}}=0$ for $X_{ij}$ has a 
solution as $X_{ij}=x_s\delta_{ij}$ with 
\begin{equation}
x_s=\frac{-m \pm \sqrt{m^2-4f\lambda}}{2\lambda} .
\end{equation}
The F-flat conditions $W_{\Phi_{ij}}=W_{\phi_i}=W_{\tilde
  \phi_j}=0$ for $Y_{ij}, \chi_i$ and $\tilde \chi_j$ have 
the following solution,
\begin{equation}
f\delta_{ij} + h \chi_i \tilde \chi_j = 0, \qquad  Y_{ij}=0.
\label{sol-Y}
\end{equation}
In addition, the D-flat conditions correspond to 
$|\chi_i| = |\tilde \chi_i|$.

There is another solution,  
$\chi_i=\tilde \chi_j=0$ and $Y_{ij}=x_s\delta_{ij}$.
However, only the above solution (\ref{sol-Y}) survives at the 
IR region, as $\hat m$ and $\hat \lambda$ become to vanish as 
we will see later.
Thus, we concentrate to the solution (\ref{sol-Y}).
At any rate, the superpotential (\ref{W-1}) does not have R-symmetry, 
and there is a supersymmetric minimum.

The above aspect is the behavior of this model around 
the energy scale $\Lambda$.
Now let us study the behavior around the IR region.
We assume that the gauge coupling is around the IR fixed point, 
i.e. $\beta_\alpha \approx 0$, and that $\phi_i$ and $\tilde \phi_i$ 
have negative anomalous dimensions $\gamma_\phi$.
In addition, we assume that the physical Yukawa coupling $\hat h$ 
is driven toward IR fixed points.
The beta-function of $\hat h$ is obtained as 
\begin{equation}
\beta_{\hat h} = \hat h (\gamma_\phi +\gamma_{\tilde \phi} + \gamma_\Phi).
\end{equation}
The condition of the fixed point leads to 
$2 \gamma_\phi + \gamma_\Phi =0$.
Since $\gamma_\phi < 0$, we obtain a positive anomalous dimension 
for $\Phi_{ij}$.
Then, physical couplings, $\hat f$, $\hat m$ and 
$\hat \lambda$, are suppressed exponentially 
toward the IR direction as 
\begin{eqnarray}
\hat f(\mu) &=& \left( \frac{\mu}{\Lambda} \right)^{\gamma_\Phi}
\hat f(\Lambda), \qquad   
\hat m(\mu) = \left( \frac{\mu}{\Lambda} \right)^{2 \gamma_\Phi}
\hat m(\Lambda), \nonumber \\   
\hat \lambda(\mu) &=& \left( \frac{\mu}{\Lambda} \right)^{3\gamma_\Phi}
\hat \lambda(\Lambda).
\end{eqnarray}
Thus, the mass parameter $\hat m$ and 3-point coupling $\hat \lambda$ 
are suppressed faster than $\hat f$.
If we neglect $\hat m$ and $\hat \lambda$ but not 
$\hat f$, the above superpotential becomes 
the superpotential of the ISS model, and 
there is a SUSY breaking minimum around $\Phi_{ij}=0$ because of 
the rank condition.

Let us see more explicitly.
We concentrate ourselves to the potential of 
the fields $X_{ij}$, because $X_{ij}$ 
contribute to SUSY breaking in the ISS model.
Furthermore, we consider their overall direction, 
i.e. $X_{ij}=x\delta_{ij}$, and we use the 
canonically normalized basis, $\hat x$.
Then, the above superpotential (\ref{W-1}) leads to the 
following scalar potential,
\begin{equation}
V_{\rm SUSY} = (N_f -N)|\hat f + \hat m \hat x +\hat \lambda \hat x^2|^2 .
\end{equation}
In addition, around $\hat x=0$, SUSY is broken and 
that generates one-loop effective potential of $\hat x$.
Around $\hat x=0$, the mass term $m_x^2|\hat x|^2$ 
in the one-loop effective potential would be important.
Hence, we analyze the potential, $V= V_{\rm SUSY} + m_x^2|\hat x|^2$, 
and we use $m_x^2$, which has been calculated in
\cite{Intriligator:2006dd}, i.e.
\begin{equation}
m_x^2=\frac{\hat h^3 \hat f}{8 \pi^2}N(N_f - N) (\log 4 -1 ).
\end{equation}
Note that $m_x^2$ is suppressed toward the IR region like $\hat f$.
We consider only the real part of $\hat x$.
The stationary condition $\partial_{\hat x} V=$ is written as 
\begin{equation}
(\hat f + \hat m \hat x +\hat \lambda \hat x^2)(\hat m + 2\hat \lambda
\hat x) + m_x^2\hat x=0.
\label{st-cod}
\end{equation}
At a high energy scale corresponding to $Z_\Phi=O(1)$, 
we have $|\hat f|, |\hat m|^2 \gg m^2_x$, because 
$m_x^2$ is smaller than $\hat f$ by a loop factor.
The potential and the stationary condition are 
controlled by $|\hat f|, |\hat m|^2$, $\hat \lambda$, 
but not $m_x$.
Thus, there is no (SUSY breaking) minimum around $x=0$, 
but we have a supersymmetric minimum 
\begin{equation}
\hat x_s =\frac{-\hat m \pm 
\sqrt{\hat m^2-4\hat f\hat \lambda}}{2\hat \lambda} .
\label{susy-min}
\end{equation}
However, toward the IR direction, 
$\hat m^2$ becomes suppressed faster than $m_x^2$.
Then, the couplings $\hat f$ and $m_x^2$ are important 
in the potential.
Around $\hat x=0$, the stationary condition (\ref{st-cod}) becomes 
\begin{equation}
\hat f \hat m + m_x^2\hat x + \cdots = 0,
\end{equation}
that is, the stationary condition is satisfied 
with 
\begin{equation}
\hat x_{sb} \approx - \frac{\hat f \hat m}{m_x^2}.
\end{equation}
At this point, SUSY is broken, 
and this point becomes close to $\hat x_{sb} =0$  toward the IR.
Around $\hat x=0$, the size of mass is estimated by 
$m_x$, because the other terms are suppressed.
Hence, the SUSY breaking metastable vacuum 
corresponding to $\hat x \sim 0$ 
appears at the IR energy scale, where $\hat m^2 \ll m_x^2$.
Moreover, the previous SUSY vacuum (\ref{susy-min}) 
moves to a point far away from the origin $\hat x=0$, 
because it behaves like
\begin{equation}
\hat x_s =\frac{-\hat m \pm \sqrt{\hat m^2-4\hat f\hat \lambda}}{2\hat \lambda} 
\sim \left( \frac{\Lambda}{\mu} \right) ^{\gamma_\Phi}.
\end{equation}

Both breaking scales of the $SU(N)$ gauge symmetry and supersymmetry 
at the metastable SUSY breaking point $\hat x =0$ are 
determined by $O(\hat f(\mu))$.
Thus, such an energy scale is estimated as 
$\mu_{IR}^2 \sim \hat f(\mu_{IR})$, i.e. 
\begin{equation}
\mu_{IR} \sim \left( \frac{\hat f(\Lambda)}{\Lambda^{\gamma_\Phi}} 
\right)^{1/(2-\gamma_\Phi)},
\end{equation}
and at this energy scale conformal renormalization group flow 
is terminated.

So far, we have assumed that the mass term of 
$\phi_i$ and $\tilde \phi_i$, $m_\phi \phi_i \tilde \phi_i$ vanishes.
Here, we comment on the case with such terms. 
The physical mass $\hat m_\phi$ becomes enhanced as 
\begin{equation}
\hat m_\phi(\mu) = \left( \frac{\mu}{\Lambda} \right)^{2 \gamma_\phi} 
 \hat m_\phi(\Lambda),
\end{equation}
because of the negative anomalous dimension $\gamma_\phi$.
At $\mu \sim \hat m_\phi(\mu)$, the matter fields $\phi_i$ $\tilde \phi_i$
decouple and this theory removes away from the conformal window.
Thus, if $\hat m_\phi(\mu) > \mu_{IR}$, the conformal renormalization 
group flow is terminated at $\mu_D \sim  \hat m_\phi(\mu_D)= 
(\mu_D/\Lambda)^{2\gamma_\phi}\hat m_\phi(\Lambda)$.

We have studied the scenario that 
conformal dynamics leads to metastable SUSY breaking 
vacua.
As an illustrating example of our idea, 
we have used the simple model.
Our scenario could be realized by other models.

\section{5D model}

There would be an AdS dual to our conformal scenario.
Indeed, we can construct simply various models 
within the framework of 5D orbifold theory.
Renormalization group flows in the 4D theory correspond to 
exponential profiles of zero modes like $e^{-c_i Ry}$, where 
$R$ is the radius of the fifth dimension,\footnote{
We assume that the radion is stabilized.} 
$y$ is the coordinate for the extra dimension, i.e. $y=[0,\pi]$ and 
$c_i$ is a constant.
The parameter $c_i$ corresponds to anomalous dimension in the 4D theory,
and each field would have a different constant $c_i$.
In 4D theory, values of anomalous dimensions are constrained 
by concrete 4D dynamics.
However, constants $c_i$ do not have such strong constraints, 
although they would correspond to some charges.
Hence, 5D models would have a rich structure 
and one could make model building rather simply.
Here we show a simple 5D model.
We consider the 5D theory, whose 5-th dimension is 
compactified on $S^1/Z_2$. 
Two fixed points on $S^1/Z_2$ correspond to $y=0$ and $y=\pi$.
We introduce three bulk fields $X$, $\phi_1$, $\phi_2$.
They correspond to chiral multiplets of 
bulk hyper-multiplets and zero modes of their partners in hyper-multiplets 
$X^c$, $\phi_1^c$, $\phi_2^c$
are projected out by the $Z_2$ orbifold projection.
We assume that zero mode profiles of $X$, $\phi_1$ and $\phi_2$ 
behave along the $y$ direction as $e^{-c_X R y}$, $e^{-c_1 Ry}$ and 
$e^{-c_2 Ry}$, respectively.
We integrate $y$ and obtain their kinetic term coefficients $Y_i$ of 
4D effective theory, that is, the field corresponding to 
the zero mode profile  $e^{-c_i Ry}$ has the following 
kinetic term coefficient \cite{Gherghetta:2000qt,Marti:2001iw}
\begin{equation}
Y_i =\frac{1}{c_i}\left( 1 - e^{-2c_i\pi R} \right).
\end{equation}
In the limit $c_i \rightarrow 0$, $Y_i$ becomes $2\pi R$.
Their superpotential is not allowed in the bulk, 
but is allowed on the boundary.

Suppose that the following superpotential is allowed 
only on the $y=\pi$ boundary,
\begin{equation}
\int dy \delta (y-\pi) W^{(\pi)},
\end{equation}
\begin{eqnarray}
W^{(\pi)} & = & fe^{-c_X Ry}X + me^{-2c_X Ry}X^2 + 
h  e^{-3c_X Ry}X^3  \nonumber \\
 & & +m_{12} e^{-(c_1+c_2) Ry}\phi_1 \phi_2
+ m_2 e^{-2c_2 Ry}\phi_2^2  \\
 & & +\sum_{i,j} h_{ij} e^{-(c_X+c_i+c_j) Ry} X \phi_i \phi_j.
\nonumber
\end{eqnarray}
Here we have assumed extra $Z_2$ symmetry, under which 
$X$ has the even $Z_2$ charge and $\phi_1$ 
and $\phi_2$ have the odd $Z_2$ charge.
That allows the mass term $m_{11}\phi^2$, but 
we have assumed it vanishes by the same reason as 
why we did not add the mass term $m_{ij} \phi_i \tilde \phi_j$ 
in the superpotential (\ref{W-1}).
We assume that $f \approx m^2 \approx m_{12}^2 \approx m_2^2$ 
and $h,h_{ij} =O(1)$.
We take  
\begin{equation}
c_1 = 0, \qquad c_2=c_X,
\end{equation}
and $c_X >0$ with $c_X\pi R =O(1)$ and 
$e^{-c_X\pi R }\ll 1$.
The 4D superpotential $\hat W$ becomes 
\begin{equation}
\hat W = e^{-c_X\pi R}(fX+m_{12}\phi_1\phi_2 + h_{11}X\phi_1^2)
 + e^{-2c_X\pi R}\Delta W.
\label{W-4D}
\end{equation}
When we neglect $\Delta W$, the superpotential $W$ 
corresponds to the O'Raifeartaigh model \cite{O'Raifeartaigh:1975pr}, 
that is, SUSY is broken.
Such a minimum is metastable and there is a SUSY minimum, 
when we take into account $\Delta W$ \cite{Abe:2007ax}.
The O'Raifeartaigh model with the following superpotential,
\begin{equation}
W_{0} = \hat f X + \hat m_{12} \phi_1 \phi_2 
+ \hat h_{11} X \phi_1^2,
\end{equation}
leads to the SUSY breaking minimum of scalar potential 
$V=|\hat f|^2$ at $\phi_1=\phi_2=0$ and arbitrary $X$, 
that is, it has the pseudo-flat direction.
One-loop effects lift up this pseudo-flat direction, 
and the field $X$ has the mass $m_X$, 
\begin{equation}
m_X^2 = O\left( \frac{1}{4\pi^2} 
\frac{\hat f^2 \hat h^4_{11}}{\hat m_{12}^2}\right),
\end{equation}
around $X=0$.
In the case with $h_{11}=O(1)$ in the superpotential (\ref{W-4D}), 
we would have a rather small mass $m_X$ by the suppression factor 
$e^{-c_X \pi R}$.
To have larger mass $m_X$, we can assume the following superpotential 
$W^{(0)}$ at $y=0$ as 
\begin{equation}
\int dy \delta(y) W^{(0)}= \int dy \delta(y) h^{(0)}_{11}X\phi_1^2.
\end{equation}
In this case, the 4D superpotential becomes
\begin{equation}
\hat W = (h^{(0)}_{11}-h_{11}e^{-C_X \pi R})X\phi_1^2  + 
e^{-c_X\pi R}(fX+m_{12}\phi_1\phi_2 )
 + e^{-2c_X\pi R}\Delta W.
\end{equation}
This leads to the metastable SUSY breaking minimum around $X=0$ 
and the field $X$ can have a larger mass around $X=0$ than the 
previous model, because the coupling $h^{(0)}_{11}$ has no 
suppression factor like $e^{-c_X\pi R}$.
The SUSY breaking source $F^X$ is quasi-localized around $y=0$.

We can construct more various models for approximately 
R-symmetric superpotential with metastable SUSY breaking 
vacua in 5D theory.
We would discuss such models elsewhere.

\section{Conclusion and discussion}

We have studied the scenario that conformal dynamics leads to 
approximately R-symmetric superpotential with a metastable 
SUSY breaking vacuum.
We have shown a simple model to realize our scenario.
We can make 5D models with the same behavior.
Since in our 4D scenario, metastable SUSY breaking vacua 
are realized by conformal dynamics, 
such a SUSY breaking source would be sequestered from the 
visible sector by conformal dynamics.

In our scenario, at a high energy scale, there would be 
only SUSY minimum and at low energy metastable 
SUSY breaking vacuum would appear.
To realize the initial condition such that 
a metastable SUSY breaking is favored at a high energy scale,
finite temperature effects would be important, 
because finite temperature effects might 
favor a metastable SUSY breaking vacuum \cite{Abel:2006cr}.

\subsection*{Acknowledgement}
H.~A.\/ and T.~K.\/ are supported in part by the
Grand-in-Aid for Scientific Research \#182496 
and \#17540251, respectively.
T.~K.\/ is also supported in part by 
the Grant-in-Aid for
the 21st Century COE ``The Center for Diversity and
Universality in Physics'' from the Ministry of Education, Culture,
Sports, Science and Technology of Japan.

\end{document}